\documentclass[doublecol]{epl2} 

\title{Spin-glasses in optical cavity}

\author{Chiu Fan Lee\inst{1}\footnote{Email: C.Lee1@physics.ox.ac.uk} \and 
Neil F. Johnson\inst{2}}
\shortauthor{C. F. Lee and N. F. Johnson}

\institute{                    
  \inst{1} Physics Department, Clarendon Laboratory, 
Oxford University, Parks Road, Oxford OX1 3PU, UK\\
  \inst{2} Physics Department, University of Miami, Coral Gables, Florida 
FL 33124, USA
}


\abstract{
Recent advances in nanofabrication and optical control have garnered 
tremendous interest in multi-qubit-cavity systems.
Here we analyze a spin-glass version of such a nanostructure, solving 
analytically for the phase diagrams in both the matter and radiation
subsystems in the replica symmetric regime. Interestingly, the resulting 
phase transitions turn out to be tunable simply by varying the
matter-radiation coupling strength.}

\begin{document}

\maketitle

Atomic physics, nanostructure materials science and quantum optics have 
recently made remarkable advances in the 
fabrication and manipulation of matter-radiation systems \cite{books}. 
The energy gap   in a semiconductor quantum
dot \cite{QD} can be engineered by varying the dot size and choice of 
materials. For
example, vanishingly small optical gaps could be obtained using InAs/GaSb 
or
HgTe/CdTe quantum dots, while vanishingly small inter-subband gaps can be 
obtained by
increasing the dot's size  \cite{QD}. 
Hence tailor-made 
two-level
`qubit' (i.e. quantum bit) systems 
\cite{nielsen} can be built on the length-scale of $10^2-10^3$ Angstroms, 
and with various geometric shapes, using 
a range of III-V and II-VI semiconductors \cite{QD}.
In addition, experimental control
of the qubit-cavity coupling $\lambda$ has already been demonstrated 
\cite{NatQED,review1, review2} for 
quantum dots coupled to photonic band-gap
defect modes \cite{PBG}, as well as for atomic and superconducting qubit
systems. A qubit-qubit interaction can arise, for example, from the 
electrostatic 
inter-dot dipole-dipole interaction between excitons and/or conduction 
electrons, and can be engineered by 
adjusting the
quantum dots' size, shape, separation, orientation and the background 
electrostatic 
screening. 
The interaction's anisotropy can be engineered by choosing 
asymmetric dot
shapes. Disorder in the qubit-qubit interactions can be introduced by 
varying
the individual dot positions during fabrication, or will arise naturally 
for
self-assembled dots \cite{QD,jopa}. All the 
pieces are therefore
in place for engineering all-optical realizations of condensed
matter spin-based systems.

Given these exciting developments in multi-qubit-cavity nanostructure, we 
study here the effect of a photon field 
on a set of qubits with disordered interaction, which can be viewed as a 
spin-glass \cite{EA, Sherr}. 
More specifically, we provide an analytic analysis of the phase behaviour 
with respect to temperature, spin-spin coupling strength and photon-spin 
coupling strength variations. We
find that phase
transition phenomena arise within both the spin (i.e. matter) and boson 
(i.e. photon) subsystems. 
With the current technology, an order of $10^3$ quantum dots can be 
embedded in a cavity structure \cite{review1}, the phase transition should 
therefore be a prominent feature of the system. Also importantly from an 
experimental point of 
view, the resulting phase diagrams can be
explored within a given nanostructure array by 
varying the qubit-cavity coupling strength $\lambda$.
Furthermore,  single quantum dot readout is currently under intense 
development (e.g., see \cite{QDmeasurement}). This opens up the possibility 
of studying the local feature of the system, which would be extremely 
helpful in understanding spin glasses.

Our main results follow from solving analytically, in the replica symmetric 
regime, a generalised version of the Dicke model \cite{Dic54, CMP}. The 
Dicke model was originally developed to describe the radiative decay
of a gas of two-level systems \cite{Dic54}, the superradiance-subradiance 
phase transition was later discovered \cite{WH73}. Debates then ensued as 
to whether the phase transition is physical, due to the constraint of the 
Thomas-Reiche-Kuhn (TRK) sum rule  (e.g., see \cite{debate}). However 
Keeling has recently provided a convincing argument for the physical 
existence of the Dicke phase transition by analysing the full atom-photon 
hamiltonian \cite{Keeling07}. 
Furthermore, in contrast to the debates concerning atom-photon 
interactions, our primary
concern here is solid-state systems such as quantum dots where there are 
usually many electrons. As such, the dipole strengths can be re-distributed 
among the different energy levels, and hence the constraint from the TRK 
sum rule can be drastically modified for the lowest two energy levels 
\footnote{Interestingly, the borrowing of dipole strength from other energy 
levels seem to be exploited by some photosynthetic systems in nature as 
discussed by Hu {\it et al.},  {\em Quart. Rev. Biophys.} {\bf 35} (2002) 
1.}. For these reasons, we believe that it is indeed legitimate for us to 
employ here the original Dicke hamiltonian, supplemented with a spin-spin 
coupling term to model our qubit-qubit interaction. More specifically, our 
hamiltonian is of the form:
\begin{equation} 
H=a^\dag a +
  \sum_{j=1}^{N}  \frac{\lambda}{2 \sqrt{N}}
  (a +a^\dag)(\sigma^+_j+
  \sigma^-_j) +
\sum_{j}\frac{\epsilon}{2}\sigma^Z_j +H_{SS}  
\end{equation}
where the operators $a, a^\dag$ and $\sigma^{\pm}_j, \sigma^Z_j$ correspond 
to the photon field and quantum dot $j$ respectively, and $H_{SS}$ is the 
spin-spin coupling term. 

We now make the following
assumptions:
\begin{enumerate}
\item
$H_{SS}$ is of the form $\sum_{i<j}J_{ij}\sigma^X_i \sigma^X_j$.
\item
$\epsilon \ll \min\{1,\lambda / \sqrt{N}, \{J_{ij}\} \}$
\end{enumerate}
As discussed in \cite{review2}, there are many ways to realize the first 
assumption. For example, each quantum dot can be engineered to have an 
elongated form along the $x$-direction, by biasing the growth process along 
this direction. Applying an electric field along $x$, will then create 
large permanent
dipole moments in that direction. One can use undoped dots, in which case 
the dipole results from the exciton, or doped dots, in which case the 
dipole originates from the conduction-subband electron biased along $x$. 
The first assumption also implies that the coupling in the $x$ direction  
overwhelms the (multipole) coupling in the $y$ and $z$ directions. 
The second assumption requires a 
negligible energy gap between the two levels of the quantum dot. This could 
be achieved using HgTe/CdTe quantum dots in order to reduce $\epsilon$, 
and/or by engineering strong dot-dot interactions $\{J_{ij}\}$ and a strong 
cavity-dot optical coupling $\lambda$. 

We would like to note that we have studied photon-spin-glasses systems in a 
more general 
setting in \cite{Tim} where the phase diagram is analysed by numerically 
solving a set of self-consistent equations deduced using the Trotter-Suzuki 
method \cite{Trotter}. Although the hamiltonian employed here is more 
restrictive by comparison, it
has the  virtue of being amenable for analytical treatment as we shall see 
below. 

For self-assembled dots, the dot-dot coupling terms $J_{ij}$ will have an 
inherent disorder. Alternatively, such disorder can be built in during 
growth by varying the dot-dot separations. 
As a general model for disorderness, we make the assumption 
that the distribution of $J_{ij}$'s is Gaussian:  
\begin{equation}
P(J_{ij})=\frac{1}{J}\sqrt{\frac{N}{2 \pi}} \exp \left\{
-\frac{N}{2J^2} \left( J_{ij} -\frac{J_0}{N} \right)^2 \right\}
\end{equation}
with $J_0$ and $J$ representing the mean and standard deviation of the 
probability
distribution. We note that negative $J_{ij}$ is 
experimentally feasible given the 
multipole (e.g. dipole-dipole) nature of the qubit-qubit interactions.

We now 
introduce the Glauber coherent states $| \alpha \rangle $ \cite{Glauber}, 
which have
the following properties:
$a|\alpha \rangle = \alpha | \alpha \rangle$,
  $\langle \alpha | a^\dag = \langle \alpha | \alpha^*$, and $\int \frac{d  
{\rm Re} (\alpha) d {\rm Im} (\alpha) }{\pi}  |\alpha \rangle \langle  
\alpha| =1$. In terms of this basis, the canonical partition function can 
be written as:
\begin{equation} Z(N,T)=\sum_{\bf s} \int \frac{d {\rm Re} (\alpha) d {\rm 
Im} (\alpha) }{\pi}  \langle \alpha| {\rm Tr} e^{-\beta H} | \alpha 
\rangle \ .\end{equation} 
We adopt the same assumptions 
as in \cite{WH73}: (i)
the
order of the double limit in  
$\lim_{N \rightarrow \infty} \lim_{R \rightarrow \infty} \sum_{r=0}^R 
\frac{(-\beta H_N)^r}{r!}$ can be interchanged, and (ii) $a/\sqrt{N}$ and 
$a^\dag/
\sqrt{N}$ exist as $N \rightarrow \infty$.  With these assumptions, we 
rewriting $Z(N,T)$ in the following form:
\begin{equation}
Z(N,T) = \int \frac{d^2 \alpha}{\pi} e^{-\beta|\alpha|^2}{\rm Tr} 
e^{-\beta H'}
\end{equation}
where
\begin{equation} 
H' = 
  \sum_{j} \frac{2\lambda {\rm Re}(\alpha)}{\sqrt{N}}
 \sigma^X_j - \sum_{i<j}
   J_{ij} \sigma^X_i  \sigma^X_j 
 \ \ .
\end{equation} 
It should now be clear why we made the assumptions concerning the form of 
our hamiltonian -- {\it all the terms in $H'$ are commutative and so we can 
integrate out $\alpha$ analytically}. Again, if the terms are not 
commutative, numerical methods will have to be used \cite{Tim}. 

Performing the Gaussian integral with respect to $\alpha$, we obtain
\begin{equation}
Z=\frac{1}{\beta} {\rm Tr} \exp \left\{-
\beta \sum_{i<j}\left(-J_{ij}-\frac{2\lambda^2}{N}\right) \sigma^X_i   
\sigma^X_{j} + \beta \lambda^2
\right\}\ \ .
\end{equation}
Hence the problem can be mapped onto the traditional spin-glass  
hamiltonian {\em if} we make the transformation $J_0\rightarrow 
\tilde{J}_0$ 
in the $J_{ij}$ probability  
distribution, where  
$\tilde{J}_0=(J_0+2\lambda^2)$. In other words, the phase diagram for the 
matter system (e.g., a nanostructure array) is equivalent to the usual 
spin-glass one \cite{Sherr} (c.f. Fig. 1).

\begin{figure}
\caption{
Phase diagram of the matter (e.g. quantum dot) subsystem within a
multi-qubit-cavity such as a disordered quantum dot array coupled to 
an optical
cavity mode. The scaled parameter $\tilde{J}_0$ can be varied simply by 
changing
the qubit-cavity coupling strength $\lambda$ (see text). $T$ is the 
temperature.}
\begin{center}
\includegraphics[scale=0.5]{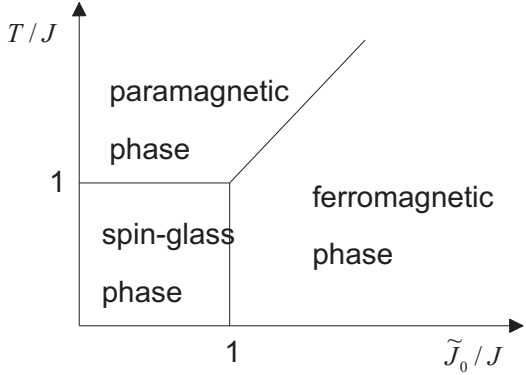}
\end{center}
\end{figure}
\begin{figure}
\caption{
Phase diagram showing the superradiant-subradiant
transition in the optical subsystem. 
In the absence of spin-spin
interactions, the system is always superradiant due to the two-level 
system's
negligible energy gap $\epsilon$.}
\begin{center}
\includegraphics[scale=0.5]{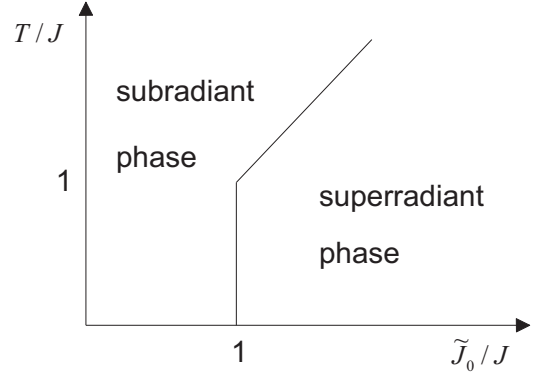}
\end{center}
\end{figure}

We now turn to consider the superradiant and subradiant photon
states in the optical subsystem. 
We recall that the order parameter for the subradiant-superradiant phase 
separation is defined to be \cite{WH73}:
\begin{equation}
\theta \equiv \left\langle \frac{a^\dag a}{N} \right\rangle 
\end{equation}
In view of this, we insert a factor $h$ in the exponent 
$-\beta |  
\alpha|^2$ in $Z$, i.e.,
\begin{equation}
Z = \int \frac{d^2 \alpha}{\pi} e^{-h\beta|\alpha|^2}{\rm Tr} e^{-\beta H'} 
\end{equation} 
so that
\begin{equation}
\theta = \left.  
\frac{1}{ N}\frac{\partial [F]}{\partial h}\right|_{h=1}\ .
\end{equation}
We now integrate out $\alpha$ in $Z$ and we obtain:
\begin{equation}
Z=\frac{1}{\beta} {\rm Tr} \exp \left\{-
\beta \sum_{i<j}\left(-J_{ij}-\underline{\frac{2\lambda^2}{hN}}\right) 
\sigma^X_i   
\sigma^X_{j} + \beta \lambda^2
\right\}\ .
\end{equation}
Namely, the only modification is the extra $h$ in the denominator in the 
underlined term. 

We now recall the definition of the free energy for spin glass systems 
\cite{Sherr, MPV87}: 
\begin{equation}
[F]=-\frac{1}{\beta}[\log Z]=-\frac{1}{\beta}\int d[J]P(J_{ij}) \log Z
\end{equation}
where $d[J] = \prod_{ij}dJ_{ij}$ and we employ the replica method to 
approximate $[\log Z]$ as $\lim_{n \rightarrow 0} 
\frac{[Z^n]-1}{n}$.
Going through the  
standard procedure of integrating out the quenched disorder in $J$s 
\cite{MPV87}, we obtain:
\begin{equation}
[Z^n]=e^{N\beta^2J^2n/4}\int d[q] d[m] e^\Omega 
\end{equation}
where $d[q] \equiv \prod_{u<v} dq_{uv}$, $d[m] \equiv \prod_u dm_u$ 
and
\begin{eqnarray}
\Omega &=& -\frac{N\beta^2 J^2}{2} \sum_{u \neq v} q_{uv}^2-
\frac{N \beta \tilde{J}_0}{2} \sum_u m_u^2 
\nonumber
\\
&&+\frac{N\beta^2 J^2n}{4}+N\log {\rm Tr} e^L 
\end{eqnarray}
with 
\begin{equation}
L=\beta^2 J^2 \sum_{u<v}q_{uv} \sigma^u \sigma^v + \beta \sum_u
\tilde{J}_0 m_u \sigma^u
\end{equation}
where $u,v$ are the replica indices and the $X$-superscripts in the 
$\sigma$s are dropped for clarity. Note that $\tilde{J_0}$ is now $(J_0+2 
\lambda^2/h)$ and so we obtain for $\partial [Z^n]/\partial h$ the 
following expression:
\begin{eqnarray}
e^{N\beta^2J^2n/4} \times 
\nonumber
\\
\int d[q] d[m]  
\left\{
\frac{2N\beta \lambda^2}{h^2} \sum_u (-m_u+2 \langle\sigma^u\rangle_L ) m_u  
\right\}
e^\Omega
\end{eqnarray}
where $\langle\sigma^u\rangle_L \equiv\frac{{\rm Tr} \sigma^u e^L}{{\rm Tr} 
e^L}$.
It follows that 
\begin{eqnarray}
\theta &=& \left.  
\frac{1}{ N}\frac{\partial [F]}{\partial h}\right|_{h=1}
=\left. -\frac{1}{n\beta N} \frac{\partial [Z^n]}{\partial h}\right|_{h=1}
\nonumber
\\
&=&\frac{2\lambda^2}{n} e^{N\beta^2J^2n/4} \times
\nonumber
\\
&&
\int d[q] d[m] 
\left\{
\sum_u (2 \langle\sigma^u\rangle_L -m_u) m_u \right\}
e^\Omega\  .
\end{eqnarray}
Since $\Omega$ is proportional to $N$, we can evaluate the  
integral by steepest-descent. In the thermodynamic limit $N  
\rightarrow \infty$, we find that
\begin{equation}
\theta={\rm const.} \times \frac{1}{n}
\sum_u (2 \langle\sigma^u\rangle_L -m_u) m_u e^{\Omega_{\rm max}}\ \ .
\end{equation}
Since $\Omega$ is optimized when
\begin{equation}
q_{uv} = \frac{1}{\beta^2 J^2} \frac{\partial}{\partial q_{uv}}
\log {\rm Tr} e^L\ = \ \langle \sigma^u \sigma^v \rangle_L 
\end{equation}
and
\begin{equation}
m_u = \frac{1}{\beta \tilde{J}_0} \frac{\partial}{\partial m_{u}}
\log {\rm Tr} e^L \ = \ \langle \sigma^u \rangle_L\ \ ,
\end{equation}
we have $\theta \propto \frac{1}{n} \sum_u m_u^2$. Therefore,
in the replica-symmetric case, $\theta \propto m^2$ and one can then derive  
the superradiant-subradiant phase diagram as shown in Fig. 2, outlining 
the region where $m \neq 0$. 
Finally we note, by observation of Eq. (7), that spin-glass behaviour can 
{\em also}
arise in a multi-qubit-cavity system with
disorder in both 
$\{\lambda\}$ and
$\{J\}$, as long as some of the
$J$'s are negative.

In conclusion, we have analyzed a novel optical realization of a 
Hamiltonian system which is of great interest within the 
condensed matter community, and have deduced analytically the corresponding 
phase diagrams. 
In contrast to traditional realisations using magnetic 
solids,
the phase transitions in this system can be explored simply by changing the 
matter-radiation coupling strength.

\acknowledgments
CFL thanks the Glasstone Trust (Oxford) and Jesus College (Oxford) for 
financial  
support. We are grateful to Alexandra
Olaya-Castro, Luis Quiroga and Tim Jarrett for useful 
discussions.

\end{document}